\documentclass[12pt,a4paper]{article}
\usepackage[latin1]{inputenc}
\usepackage[table]{xcolor}

\usepackage{booktabs}

\usepackage{graphicx}

\usepackage{amsfonts}
\usepackage{amsmath}
\usepackage{amssymb}
\usepackage{theorem}
\usepackage{upgreek}

\theoremstyle{change}
\theorembodyfont{\slshape}

\theorembodyfont{\upshape\small}

\theorembodyfont{\ttfamily}

\newcommand{\ba}{\begin{equation}}
\newcommand{\ea}{\end{equation}}

\usepackage{dsfont}

\newcommand{\poi}{\textup{Poi}}

\newcommand{\bbn}{\mathbb{N}}
\newcommand{\bbr}{\mathbb{R}}

\newcommand{\iid}{i.\,i.\,d.}
\newcommand{\ie}{i.\,e., }
\newcommand{\eg}{e.\,g., }

\usepackage{natbib}

\usepackage{hyperref}

\begin{document}



\parindent 0cm

\title{Control Charts for Poisson Counts based on the Stein--Chen Identity}
\author{
Christian H.\ Wei\ss\thanks{
Helmut Schmidt University, Department of Mathematics and Statistics, Hamburg, Germany.}\ \thanks{Corresponding author. E-Mail: \href{mailto:weissc@hsu-hh.de}{\nolinkurl{weissc@hsu-hh.de}}. ORCID: \href{https://orcid.org/0000-0001-8739-6631}{0000-0001-8739-6631}}
}

\maketitle

\begin{abstract}
\noindent
If monitoring Poisson count data for a possible mean shift (while the Poisson distribution is preserved), then the ordinary Poisson exponentially weighted moving-average (EWMA) control chart proved to be a good solution. In practice, however, mean shifts might occur in combination with further changes in the distribution family. Or due to a misspecification during Phase-I analysis, the Poisson assumption might not be appropriate at all. In such cases, the ordinary EWMA chart might not perform satisfactorily. Therefore, two novel classes of generalized EWMA charts are proposed, which utilize the so-called Stein--Chen identity and are thus sensitive to further distributional changes than just sole mean shifts. Their average run length (ARL) performance is investigated with simulations, where it becomes clear that especially the class of so-called ``ABC-EWMA charts'' shows an appealing ARL performance. The practical application of the novel Stein--Chen EWMA charts is illustrated with an application to count data from semiconductor manufacturing. 

\end{abstract}

\section{Introduction}
\label{Introduction}
Count data are often observed in quality-related contexts, such as counts of non-conformities on manufactured items, counts of customer complaints in service industries, and counts of infections or hospital admissions in health surveillance. For an early detection of possible quality problems, (attributes) control charts are commonly applied, where the sequence of incoming (samples of) counts is used for an online computation of statistics, which are then plotted on the chart until an alarm is triggered. The latter happens if the plotted statistic violates the specified control limits (CLs), which are determined based on a given in-control model (\ie a stochastic model for the count process if this operates under stable conditions). 
In practice, the in-control model is identified based on historical data (``Phase-I analysis''), and the subsequent prospective online monitoring of novel data is called the Phase-II application of the control chart. 
An alarm during Phase~II is understood as an indication of the possible occurrence of an out-of-control situation, whereas the process is considered in control otherwise. Obviously, the CLs should be chosen such that the control chart does not trigger a false alarm (\ie while the process is in control) for as long as possible, whereas a true alarm for an out-of-control process should arrive as early as possible. Therefore, the design of the control chart is typically chosen with respect to the resulting average run length (ARL) performance, \ie the mean number of plotted points until we get the first alarm in a selected scenario. For references and further details on these basic terms and concepts of statistical process control (SPC), the reader is referred to standard textbooks such as \citet{mont09,qiu14}.

\smallskip
For the sake of a concise presentation, we shall focus on a particular (and popular) type of count process $(X_t)_{t\in\bbn}$ with $\bbn=\{1,2,\ldots\}$, namely on individual univariate unbounded counts (\ie where the range is given by the full set of non-negative integers, $\bbn_0=\{0,1,\ldots\}$) following a Poisson distribution under in-control conditions. But later in Section~\ref{Conclusions}, it is also discussed how the proposed approaches could be extended to different types of count distributions, or even to continuously distributed variables data. Furthermore, we assume $(X_t)_{t\in\bbn}$ to be serially independent both under in-control and out-of-control conditions, being aware that the case of autocorrelated counts needs to be investigated in future research, see Section~\ref{Conclusions}. 
Hence, from now on, let $(X_t)_{t\in\bbn}$ be independent and identically distributed (\iid) according to $\poi(\mu)$, the Poisson distribution with mean~$\mu>0$, where $\mu=\mu_0$ in the in-control case. In most research papers (also see the recent survey article by \citet{alevizakos20}), control charts for Poisson counts are analyzed with respect to their ability to discover changes in the mean, \ie $\mu\not=\mu_0$, while the Poisson distribution is assumed to hold also under out-of-control conditions. Examples are the basic c-chart \citep[see][]{mont09,qiu14}, being a memory-less Shewhart chart where the individual counts~$X_t$ are directly plotted on a chart with specified lower and upper CL (LCL and UCL, respectively), or several types of control chart with an inherent memory, \eg types of exponentially weighted moving-average (EWMA) charts \citep[\eg][]{gan90,borror98,rakitzis15,morais18,morais20}. Here, the default way of computing EWMA statistics is given by
\ba
\label{ewmarecursion}
Z_0=\mu_0,\quad
Z_t\ =\ \lambda\cdot X_t\ +\ (1-\lambda)\cdot Z_{t-1}\quad\text{for } t=1,2,\ldots
\ea
with smoothing parameter $\lambda\in (0;1]$, which are then plotted against specified LCL and UCL. Note that \eqref{ewmarecursion} reduces to the c-chart if $\lambda=1$.

\smallskip
While the aforementioned charts have been developed and analyzed for sole mean changes (\ie the Poisson assumption is preserved), their performance is not clear under more sophisticated out-of-control scenarios (\eg if the mean and the distribution family change simultaneously) or under pure changes in distribution family (\eg if the in-control distribution was misspecified but~$\mu_0$ was identified correctly). Therefore, in this paper, we shall propose two novel classes of generalized EWMA charts for Poisson counts, which are easily adapted to a broad range of out-of-control scenarios. The key idea is to utilize a type of moment identity which uniquely characterizes the Poisson distribution, the so-called \emph{Stein--Chen identity}. It is named after Charles Stein on the one hand, who developed the general idea of characterizing a certain distribution family by an appropriate moment identity \citep[see][]{stein72,stein86}, and after L.H.Y.\ Chen on the other hand, who derived the Stein-type identity for the particular case of Poisson counts \citep[see][]{chen75}.
Their result states that $X\sim\poi(\mu)$ if and only if 
\ba
\label{SteinChen}
E\big[X\, f(X)\big]\ =\ \mu\, E\big[f(X+1)\big]
\ea
holds for all bounded functions $f:\bbn_0\to\bbr$. Actually, \eqref{SteinChen} constitutes a non-trivial statement only if~$f$ is not constant on~$\bbn_0$, and if~$f$ is not identical to zero on~$\bbn$. The Stein--Chen identity \eqref{SteinChen} has been recently applied to sophisticated moment calculations \citep[see][]{weissaleksandrov22} as well as to constructing goodness-of-fit (GoF) tests \citep[see][]{betsch22,weissetal23}. In the present paper, it is used to construct novel EWMA-type control charts for Poisson counts, see Section~\ref{Stein--Chen EWMA Charts for Poisson Counts} for the proposed approaches.
In Section~\ref{Simulation-based Performance Analyses}, we analyze their ARL performance within a simulation study. 
Section~\ref{An Illustrative Data Example} presents a real-world data example from semiconductor manufacturing to illustrate the application and possible benefits of the novel Stein--Chen EWMA charts. 
Finally, Section~\ref{Conclusions} concludes the article and outlines directions for future research.

\section{Stein--Chen EWMA Charts for Poisson Counts}
\label{Stein--Chen EWMA Charts for Poisson Counts}
As mentioned in Section~\ref{Introduction}, the Stein--Chen identity \eqref{SteinChen} was recently utilized by \citet{weissetal23} to construct GoF tests regarding the null hypothesis of Poisson counts. Their crucial idea was to choose the function~$f$ in \eqref{SteinChen} in a way that matches the alternative hypothesis under consideration. For example, if the relevant alternative are negative-binomial (NB) counts, the function~$f$ should be chosen differently for good power properties than if the alternative are zero-inflated Poisson (ZIP) counts. The intuition behind such a tailor-made choice of~$f$ is as follows. 
Within the moments $E\big[X\, f(X)\big]$ and $E\big[f(X+1)\big]$ involved in \eqref{SteinChen}, the function~$f$ acts like a weight function, which can be chosen to assign more weight on selected counts. For a ZIP-alternative, for instance, \citet{weissetal23} showed that good power is achieved by having large weight close to zero, whereas the overdispersion pattern caused by an NB-alternative (its probability mass function (PMF) looks like a flattened Poi-PMF) is better detected with large weight for large counts. Specific examples could be to use an only logarithmically increasing~$f$ for a ZIP-alternative, but a linearly increasing~$f$ for an NB-alternative; see Section~\ref{Simulation-based Performance Analyses} for further details.

\subsection{Two Classes of Generalized EWMA Statistics}
\label{Two Classes of EWMA Statistics}
In what follows, we pick up this intuition and propose unified approaches for defining generalized EWMA charts that are flexibly adapted to the anticipated out-of-control scenario. Let us denote $\mu_f(k;l_1,\ldots,l_m) := E\big[X^k\cdot f(X+l_1)\cdots f(X+l_m)\big]$ with integers $k\geq 0$ and $0\leq l_1\leq \ldots\leq l_m$. In particular, $E\big[X\, f(X)\big] = \mu_f(1;0)$ and $E\big[f(X+1)\big] = \mu_f(0;1)$. If these moments are computed with respect to the in-control distribution $\poi(\mu_0)$, we add the subscript ``0'', \ie $\mu_{f;0}(1;0)$ and $\mu_{f;0}(0;1)$, respectively. Note that depending on the choice of~$f$, a numerical computation is necessary, namely via
\ba
\label{muf_num}
\mu_{f;0}(1;0)\ \approx\ \sum_{x=0}^M x\cdot f(x)\cdot e^{-\mu_0}\,\frac{\mu_0^x}{x!}
\quad\text{and}\quad
\mu_{f;0}(0;1)\ \approx\ \sum_{x=0}^M f(x+1)\cdot e^{-\mu_0}\,\frac{\mu_0^x}{x!},
\ea
where the truncation limit~$M$ is chosen sufficiently large. In this research, we always used $M=50$. Generally, $M$ could be chosen with respect of the upper tail of $\poi(\mu_0)$, \eg such that $P(X>M)$ falls below a specified tolerance value, such as $10^{-10}$ or smaller.

\smallskip
If we expect a combination of mean change and change in distribution family, it appears reasonable to solve \eqref{SteinChen} as
\ba
\label{SCEWMA1a}
\mu\ =\ \frac{E\big[X\, f(X)\big]}{E\big[f(X+1)\big]},
\ea
and to define the EWMA scheme
\ba
\label{SCEWMA1b}
\begin{array}{@{}l@{\quad}l}
A_0 = \mu_{f;0}(1;0), & A_t\ =\ \lambda\cdot X_t\,f(X_t)\ +\ (1-\lambda)\cdot A_{t-1};
\\[1ex]
B_0 = \mu_{f;0}(0;1), & B_t\ =\ \lambda\cdot f(X_t+1)\ +\ (1-\lambda)\cdot B_{t-1};
\\[1ex]
Z_0^{\textup{AB}} = \mu_0, & \displaystyle Z_t^{\textup{AB}}\ =\ \frac{A_t}{B_t};
\qquad\text{for } t=1,2,\ldots
\end{array}
\ea
Then, the statistics $Z_t^{\textup{AB}}$ are plotted on a control chart with appropriately chosen LCL and UCL (see Section~\ref{On Approaches for Chart Design} below). For simplicity, and to distinguish it from the ordinary EWMA chart \eqref{ewmarecursion}, let us refer to \eqref{SCEWMA1b} as the ``AB-EWMA chart''. Note that \eqref{SCEWMA1b} reduces to \eqref{ewmarecursion} if we choose $f\equiv 1$ constantly. Furthermore, $B_t\not=0$ has to be ensured by a reasonable (non-negative) choice of~$f$, also recall the discussion below \eqref{SteinChen}. 

\medskip
If the anticipated out-of-control scenario does not necessarily lead to a mean change but may manifest itself solely in a change of the distribution family (\eg because the in-control model was chosen incorrectly), then it might be better to solve \eqref{SteinChen} as
\ba
\label{SCEWMA2a}
1\ =\ \frac{E\big[X\, f(X)\big]}{\mu\,E\big[f(X+1)\big]},
\ea
and to define the EWMA scheme
\ba
\label{SCEWMA2b}
\begin{array}{@{}l@{\quad}l}
A_0 = \mu_{f;0}(1;0), & A_t\ =\ \lambda\cdot X_t\,f(X_t)\ +\ (1-\lambda)\cdot A_{t-1};
\\[1ex]
B_0 = \mu_{f;0}(0;1), & B_t\ =\ \lambda\cdot f(X_t+1)\ +\ (1-\lambda)\cdot B_{t-1};
\\[1ex]
C_0 = \mu_0, & C_t\ =\ \lambda\cdot X_t\ +\ (1-\lambda)\cdot C_{t-1};
\\[1ex]
Z_0^{\textup{ABC}} = 1, & \displaystyle Z_t^{\textup{ABC}}\ =\ \frac{A_t}{B_t\,C_t};
\qquad\text{for } t=1,2,\ldots
\end{array}
\ea
We refer to \eqref{SCEWMA2b} as the ``ABC-EWMA chart'', where the statistics $Z_t^{\textup{ABC}}$ are plotted on a control chart with appropriately chosen LCL and UCL.

\subsection{On Approaches for Chart Design}
\label{On Approaches for Chart Design}
Control charts are commonly designed with respect to their ARL performance. More precisely, the CLs are chosen such that the so-called zero-state ARL \citep[see][]{knoth06} is close to a given target value ARL$_0$, such as the common textbook choice ARL$_0=370$. Here, zero-state ARL expresses that the involved run lengths are computed right from the beginning of process monitoring, and that possible process changes also happen right in the beginning (the latter is only relevant for out-of-control zero-state ARLs). But there are also ARL concepts which assume a later change point~$\tau>1$, \ie the process is in control for $t<\tau$ and turns out of control for $t\geq \tau$ (zero-state ARLs assume $\tau=1$). The conditional expected delay CED$(\tau)$ assumes that there was no false alarm for $t<\tau$, and then, it is defined as the mean number of out-of-control points until the first alarm. The limit $\tau\to\infty$ is referred to as the steady-state ARL; see \citet{knoth06} for further details. 
Such CEDs are often considered as an alternative to the zero-state ARL when evaluating the chart's out-of-control performance. 
Put simply, the zero-state ARL provides insight into the chart's performance for early process changes, and CED and steady-state ARL for late process changes. 
In the case of memory-type charts such as the EWMA charts considered here, the values of the different ARL concepts might differ, although in many cases, this difference is practically negligible (\ie the charts work similarly well for early and late changes). We shall check for possible discrepancies caused by early and late changes during our performance analyses in Section~\ref{Simulation-based Performance Analyses}.

\smallskip
In addition to the requirement that the zero-state in-control ARL is close to the target value ARL$_0$, for two-sided charts having both an UCL and LCL, one may also try to find a chart design leading to an ``unbiased'' out-of-control performance. Such unbiasedness is defined with respect to a fixed out-of-control scenario, such as $\poi(\mu)$ with $\mu\not=\mu_0$ for Poisson counts (sole mean shift). Then, the function ARL$(\mu)$ should be maximal in~$\mu_0$, and it should decrease both for increasing $\mu>\mu_0$ and decreasing $\mu<\mu_0$. For the default Poi-EWMA chart \eqref{ewmarecursion}, \citet{borror98} used symmetric CLs, \ie $\textup{UCL}-\mu_0 = \mu_0-\textup{LCL} =: L$. But as shown by \citet{morais20}, this usually leads to a (somewhat) biased ARL performance. The bias problem can be solved by choosing asymmetric CLs in an appropriate way, see Section~4 in \citet{morais20}. In the present research, however, we do not have a fixed out-of-control scenario, but we experiment with rather different distributional assumptions. For this reason, we do not search for unbiased designs here but just use symmetric CLs for simplicity. Certainly, if in some applications, one has a very specific out-of-control scenario in mind, then one should try to reach unbiasedness with respect to this scenario.

\section{Simulation-based Performance Analyses}
\label{Simulation-based Performance Analyses}
The ARL performances of the proposed Stein--Chen EWMA charts are investigated by simulations with $10^4$ replications per scenario\footnote{For the ordinary EWMA chart \eqref{ewmarecursion}, zero-state and steady-state ARLs (under Poisson assumption) can also be computed numerically by using the functions \texttt{pois.ewma.arl} and \texttt{pois.ewma.ad}, respectively, of the R-package \href{https://CRAN.R-project.org/package=spc}{\nolinkurl{spc}}. There is a good agreement to the simulated values being used here.}. Due to the strict page limit, only a few in-control and out-of-control scenarios are considered here, but more comprehensive analyses (also with further types of weight function) are recommended for future research, see Section~\ref{Conclusions}. 
As the in-control model, we use Poisson counts with either mean $\mu_0=2$ (``low counts'') or $\mu_0=5$ (``medium counts''). 
The considered out-of-control scenarios are defined as a mean shift and/or a change in the distribution family, where we restrict the set of non-Poisson distributions to two types of overdispersion model: an NB- and a ZIP-distribution. Overdispersion, which constitutes the most common type of deviation from a Poisson assumption, describes the phenomenon that the variance~$\sigma^2$ is larger than the mean~$\mu$, whereas any Poisson model exhibits equidispersion ($\sigma^2=\mu$). If using an NB-distribution, the overdispersion is caused by flattening the Poi-PMF (``regular overdispersion''), whereas a ZIP-distribution generates overdispersion by a single point mass in zero. For the sake of comparability, we parametrize~NB and~ZIP in terms of the mean $\mu>0$ and the dispersion index $I=\sigma^2/\mu>1$, which uniquely determine both distributions. For the weight function~$f$, the following three specifications are considered in accordance to the recommendations of \citet{weissetal23}: the piecewise linear weights $f(x)=|x-1|$ that led to powerful GoF-tests regarding an NB-alternative, and the root weights $f(x)=|x-1|^{1/4}$ as well as logarithmic weights $f(x)=\ln(x)$ that led to powerful GoF-tests regarding a ZIP-alternative (we use the convention $0\cdot\ln(0)=0$). Besides the resulting AB-EWMA \eqref{SCEWMA1b} and ABC-EWMA \eqref{SCEWMA2b} charts, we also evaluate the ordinary EWMA chart \eqref{ewmarecursion} as a competitor. The smoothing parameter is set at $\lambda=0.10$ to ensure a strong memory. The symmetric two-sided CLs are of the form $\mu_0\mp L$ for EWMA and AB-EWMA, and $1\mp L$ for ABC-EWMA. Here, $L$ is chosen such that the (zero-state) in-control ARL is close to the target value ARL$_0=370$.

\begin{table}[t!]
\centering
\caption{Simulated zero-state ARLs of ordinary EWMA (``EWMA''), AB-EWMA (``AB''), and ABC-EWMA (``ABC'') chart, where in-control ARL printed in bold font. Weight functions $f(x)=|x-1|$, $|x-1|^{1/4}$, and $\ln(x)$ for AB and ABC. CL~$L$ shown in italic font. In-control means $\mu_0\in\{2,5\}$, mean shifts $\mp 0.25$, and NB and ZIP with $I=5/3$. Grey highlighted parts support interpretation of blocks.}
\label{tabZeroStateARLs}

\smallskip
\resizebox{\linewidth}{!}{
\begin{tabular}{ll|r@{}r@{}r|r@{}r@{}r|r@{}r@{}r}
\toprule
$\mu_0$ & \multicolumn{1}{r|}{$\mu=$} & \multicolumn{1}{c@{}}{$\mu_0-0.25$} & \multicolumn{1}{c@{}}{$\mu_0$} & \multicolumn{1}{c|}{$\mu_0+0.25$} & \multicolumn{1}{c@{}}{$\mu_0-0.25$} & \multicolumn{1}{c@{}}{$\mu_0$} & \multicolumn{1}{c|}{$\mu_0+0.25$} & \multicolumn{1}{c@{}}{$\mu_0-0.25$} & \multicolumn{1}{c@{}}{$\mu_0$} & \multicolumn{1}{c}{$\mu_0+0.25$} \\
\midrule
 &  & \multicolumn{1}{l@{}}{EWMA} &  & \itshape 0.877 & \multicolumn{1}{l@{}}{AB} & \multicolumn{1}{c@{}}{$|x-1|$} & \itshape 1.191 & \multicolumn{1}{l@{}}{ABC} & \multicolumn{1}{c@{}}{$|x-1|$} & \itshape 0.463 \\
\midrule
2 & ZIP & 83.5 & 89.7 & 54.1 & 85.7 & 50.1 & 31.9 & 24.5 & 28.3 & 32.6 \\
 & Poi & 252.6 & \bfseries 369.1 & 106.1 & 453.6 & \bfseries 370.3 & 164.6 & 232.7 & \bfseries 370.0 & 559.8 \\
 & NB & 87.3 & 95.2 & 56.0 & 57.4 & 42.5 & 30.1 & 29.7 & 34.9 & 40.6 \\
\midrule
 &  & \multicolumn{1}{l@{}}{EWMA} &  & \itshape 1.388 & \multicolumn{1}{l@{}}{AB} & \multicolumn{1}{c@{}}{$|x-1|$} & \itshape 1.614 & \multicolumn{1}{l@{}}{ABC} & \multicolumn{1}{c@{}}{$|x-1|$} & \itshape 0.1828 \\
\midrule
5 & ZIP & 83.9 & 88.3 & 71.0 & 118.7 & 74.0 & 47.5 & 22.0 & 23.3 & 24.4 \\
 & Poi & 309.9 & \bfseries 371.4 & 185.1 & 392.0 & \bfseries 371.1 & 209.3 & 309.3 & \bfseries 370.8 & 443.0 \\
 & NB & 89.6 & 93.1 & 70.6 & 60.7 & 47.6 & 37.0 & 30.3 & 32.1 & 34.3 \\
\midrule
 &  &  &  &  & \multicolumn{1}{l@{}}{AB} & \multicolumn{1}{c@{}}{$|x-1|^{1/4}$} & \itshape 1.053 & \multicolumn{1}{l@{}}{ABC} & \multicolumn{1}{c@{}}{$|x-1|^{1/4}$} & \itshape 0.382 \\
\midrule
2 & ZIP &  &  &  & 66.2 & 37.8 & 24.5 & 18.3 & 21.2 & 24.7 \\
 & Poi & \multicolumn{3}{c|}{\cellcolor[HTML]{C0C0C0}sole mean change} & 237.3 & \bfseries 371.4 & 146.7 & 180.6 & \bfseries 370.3 & 727.5 \\
 & NB &  &  &  & 76.6 & 56.4 & 37.1 & 36.4 & 51.8 & 72.7 \\
\midrule
 &  &  &  &  & \multicolumn{1}{l@{}}{AB} & \multicolumn{1}{c@{}}{$|x-1|^{1/4}$} & \itshape 1.424 & \multicolumn{1}{l@{}}{ABC} & \multicolumn{1}{c@{}}{$|x-1|^{1/4}$} & \itshape 0.106 \\
\midrule
5 & ZIP &  & \multicolumn{1}{c@{}}{\cellcolor[HTML]{C0C0C0}pure} &  & 98.7 & 57.7 & 37.0 & 11.5 & 11.6 & 12.2 \\
 & Poi &  & \multicolumn{1}{c@{}}{\cellcolor[HTML]{C0C0C0}distrib.} &  & 310.1 & \bfseries 370.5 & 189.9 & 307.7 & \bfseries 369.4 & 454.9 \\
 & NB &  & \multicolumn{1}{c@{}}{\cellcolor[HTML]{C0C0C0}change} &  & 86.8 & 74.2 & 53.0 & 48.0 & 55.0 & 65.5 \\
\midrule
 &  &  &  &  & \multicolumn{1}{l@{}}{AB} & \multicolumn{1}{c@{}}{$\ln(x)$} & \itshape 1.089 & \multicolumn{1}{l@{}}{ABC} & \multicolumn{1}{c@{}}{$\ln(x)$} & \itshape 0.396 \\
\midrule
2 & ZIP & \multicolumn{3}{c|}{\cellcolor[HTML]{C0C0C0}zero inflation} & 69.3 & 40.2 & 25.6 & 19.3 & 22.3 & 25.5 \\
 & Poi &  &  &  & 307.1 & \bfseries 370.2 & 148.8 & 194.4 & \bfseries 369.4 & 659.9 \\
 & NB &  &  &  & 63.6 & 45.8 & 31.3 & 30.1 & 38.7 & 50.1 \\
\midrule
 &  &  &  &  & \multicolumn{1}{l@{}}{AB} & \multicolumn{1}{c@{}}{$\ln(x)$} & \itshape 1.465 & \multicolumn{1}{l@{}}{ABC} & \multicolumn{1}{c@{}}{$\ln(x)$} & \itshape 0.118 \\
\midrule
5 & ZIP &  &  &  & 102.3 & 61.0 & 38.7 & 14.4 & 14.7 & 15.3 \\
 & Poi &  &  &  & 325.2 & \bfseries 370.9 & 193.8 & 296.6 & \bfseries 371.3 & 471.5 \\
 & NB & \multicolumn{3}{c|}{\cellcolor[HTML]{C0C0C0}regular overdispersion} & 79.9 & 63.3 & 45.6 & 35.1 & 38.9 & 44.4 \\
\bottomrule
\end{tabular}}
\end{table}

\subsection{Zero-state ARL Performance}
\label{Zero-state ARL Performance}
The simulated zero-state ARLs are summarized in Table~\ref{tabZeroStateARLs}. Each $3\times 3$-block corresponds to one particular chart design, where the grey highlighted parts in Table~\ref{tabZeroStateARLs} provide a scheme for interpreting the ARLs. If we are concerned with a sole mean change (\ie while preserving the Poi-assumption), then the ordinary EWMA chart constitutes the best choice. But if a change of the distribution family happens (with or without mean shift), then there are AB- and ABC-EWMA charts with better ARL performance. In fact, the advantage offered by the AB-EWMA charts is limited; they have somewhat lower out-of-control ARLs regarding NB (especially if using $f(x)=|x-1|$) and regarding ZIP in combination with a positive mean shift (especially if $f(x)=|x-1|^{1/4}$ or $f(x)=\ln(x)$), but they do not uniquely improve across all overdispersion scenarios. But a clear advantage can be recognized for the ABC-EWMA charts, which show considerably lower out-of-control ARLs in almost any overdispersion case. Thus, the ABC-EWMA chart is clearly recommended for process monitoring in the presence of changes in the distribution family. If we anticipate a regular overdispersion pattern (caused by NB), then the best performance is achieved if using $f(x)=|x-1|$, which is plausible as its monitored statistic is related to the sample dispersion index, see \eqref{SCEWMA2a}. By contrast, zero inflation is best detected by using $f(x)=|x-1|^{1/4}$, and nearly equally well by $f(x)=\ln(x)$. The choice $f(x)=|x-1|^{1/4}$, however, leads to a relatively bad performance in the NB-case. Certainly, in practice, we often do not know the precise overdispersion pattern in advance. Thus, it might be attractive for a practitioner to have a chart that exhibits some kind of robustness towards a misspecified overdispersion pattern. From Table~\ref{tabZeroStateARLs}, we can recognize that both $f(x)=|x-1|$ and $f(x)=\ln(x)$ lead to such ``universally applicable'' ABC-EWMA charts (\ie which are sensitive to both overdispersion scenarios), where $f(x)=|x-1|$ is somewhat stronger regarding regular overdispersion, and $f(x)=\ln(x)$ regarding zero inflation. Finally, it is worth pointing out that these ABC-EWMA charts also perform very well in the case of a pure change in distribution family, as it might happen by an erroneous choice of the in-control model (also see the data example in Section~\ref{An Illustrative Data Example}).

\begin{table}[t!]
\centering
\caption{Same structure as in Table~\ref{tabZeroStateARLs}, but showing simulated CED$(100)$ values instead of zero-state ARLs.}
\label{tabCED100s}

\smallskip
\resizebox{\linewidth}{!}{
\begin{tabular}{ll|r@{}r@{}r|r@{}r@{}r|r@{}r@{}r}
\toprule
$\mu_0$ & \multicolumn{1}{r|}{$\mu=$} & \multicolumn{1}{c@{}}{$\mu_0-0.25$} & \multicolumn{1}{c@{}}{$\mu_0$} & \multicolumn{1}{c|}{$\mu_0+0.25$} & \multicolumn{1}{c@{}}{$\mu_0-0.25$} & \multicolumn{1}{c@{}}{$\mu_0$} & \multicolumn{1}{c|}{$\mu_0+0.25$} & \multicolumn{1}{c@{}}{$\mu_0-0.25$} & \multicolumn{1}{c@{}}{$\mu_0$} & \multicolumn{1}{c}{$\mu_0+0.25$} \\
\midrule
 &  & \multicolumn{1}{l@{}}{EWMA} &  & \itshape 0.877 & \multicolumn{1}{l@{}}{AB} & \multicolumn{1}{c@{}}{$|x-1|$} & \itshape 1.191 & \multicolumn{1}{l@{}}{ABC} & \multicolumn{1}{c@{}}{$|x-1|$} & \itshape 0.463 \\
\midrule
2 & ZIP & 78.8 & 90.0 & 52.3 & 84.0 & 50.4 & 31.4 & 24.5 & 28.2 & 32.5 \\
 & Poi & 246.2 & \bfseries 359.4 & 104.4 & 454.3 & \bfseries 364.2 & 164.4 & 231.9 & \bfseries 363.7 & 562.0 \\
 & NB & 83.9 & 93.2 & 54.3 & 58.0 & 41.4 & 30.4 & 29.6 & 34.7 & 39.6 \\
\midrule
 &  & \multicolumn{1}{l@{}}{EWMA} &  & \itshape 1.388 & \multicolumn{1}{l@{}}{AB} & \multicolumn{1}{c@{}}{$|x-1|$} & \itshape 1.614 & \multicolumn{1}{l@{}}{ABC} & \multicolumn{1}{c@{}}{$|x-1|$} & \itshape 0.1828 \\
\midrule
5 & ZIP & 81.0 & 87.1 & 67.4 & 118.7 & 73.3 & 48.4 & 22.5 & 23.8 & 24.8 \\
 & Poi & 304.8 & \bfseries 362.6 & 180.8 & 386.2 & \bfseries 367.0 & 208.2 & 309.5 & \bfseries 369.7 & 445.6 \\
 & NB & 85.9 & 88.8 & 69.8 & 60.8 & 48.0 & 36.5 & 30.4 & 32.5 & 33.6 \\
\midrule
 &  &  &  &  & \multicolumn{1}{l@{}}{AB} & \multicolumn{1}{c@{}}{$|x-1|^{1/4}$} & \itshape 1.053 & \multicolumn{1}{l@{}}{ABC} & \multicolumn{1}{c@{}}{$|x-1|^{1/4}$} & \itshape 0.382 \\
\midrule
2 & ZIP &  &  &  & 64.3 & 37.2 & 24.3 & 18.5 & 21.0 & 24.7 \\
 & Poi & \multicolumn{3}{c|}{\cellcolor[HTML]{C0C0C0}sole mean change} & 231.1 & \bfseries 364.9 & 146.1 & 177.6 & \bfseries 367.3 & 723.8 \\
 & NB &  &  &  & 74.7 & 54.1 & 36.8 & 36.5 & 50.4 & 71.7 \\
\midrule
 &  &  &  &  & \multicolumn{1}{l@{}}{AB} & \multicolumn{1}{c@{}}{$|x-1|^{1/4}$} & \itshape 1.424 & \multicolumn{1}{l@{}}{ABC} & \multicolumn{1}{c@{}}{$|x-1|^{1/4}$} & \itshape 0.106 \\
\midrule
5 & ZIP &  & \multicolumn{1}{c@{}}{\cellcolor[HTML]{C0C0C0}pure} &  & 97.4 & 58.2 & 36.9 & 12.7 & 13.2 & 13.6 \\
 & Poi &  & \multicolumn{1}{c@{}}{\cellcolor[HTML]{C0C0C0}distrib.} &  & 304.1 & \bfseries 363.5 & 187.6 & 310.9 & \bfseries 374.8 & 461.7 \\
 & NB &  & \multicolumn{1}{c@{}}{\cellcolor[HTML]{C0C0C0}change} &  & 84.1 & 71.1 & 52.6 & 49.5 & 56.3 & 66.8 \\
\midrule
 &  &  &  &  & \multicolumn{1}{l@{}}{AB} & \multicolumn{1}{c@{}}{$\ln(x)$} & \itshape 1.089 & \multicolumn{1}{l@{}}{ABC} & \multicolumn{1}{c@{}}{$\ln(x)$} & \itshape 0.396 \\
\midrule
2 & ZIP & \multicolumn{3}{c|}{\cellcolor[HTML]{C0C0C0}zero inflation} & 67.8 & 39.8 & 25.5 & 19.6 & 22.3 & 25.5 \\
 & Poi &  &  &  & 300.7 & \bfseries 366.7 & 148.2 & 192.1 & \bfseries 362.1 & 664.5 \\
 & NB &  &  &  & 62.7 & 45.1 & 31.6 & 29.9 & 38.4 & 48.8 \\
\midrule
 &  &  &  &  & \multicolumn{1}{l@{}}{AB} & \multicolumn{1}{c@{}}{$\ln(x)$} & \itshape 1.465 & \multicolumn{1}{l@{}}{ABC} & \multicolumn{1}{c@{}}{$\ln(x)$} & \itshape 0.118 \\
\midrule
5 & ZIP &  &  &  & 101.7 & 61.0 & 38.6 & 14.2 & 14.7 & 15.1 \\
 & Poi &  &  &  & 322.6 & \bfseries 367.1 & 191.2 & 295.6 & \bfseries 366.7 & 474.0 \\
 & NB & \multicolumn{3}{c|}{\cellcolor[HTML]{C0C0C0}regular overdispersion} & 78.5 & 61.9 & 45.0 & 34.9 & 39.2 & 44.1 \\
\bottomrule
\end{tabular}}
\end{table}

\subsection{Effect of Late Process Changes}
\label{Effect of Late Process Changes}
The results of the previous Section~\ref{Zero-state ARL Performance} refer to the case of an early process change. To analyze if these results also hold (approximately) for late changes, the zero-state ARLs have been compared to CED$(\tau)$ values with $\tau\in\{5,10,25,50,100\}$. For illustration, Table~\ref{tabCED100s} shows the CED$(100)$-counterpart to Table~\ref{tabZeroStateARLs} (very late change, close to steady-state ARL), the remaining results are available from the author upon request. Generally, it can be seen that the deviations between zero-state ARLs and CED$(100)$s are low (and those of the remaining CED$(\tau)$s are in between these extreme cases), so it does not matter in practice if the process change occurs early or late. It seems that the Stein--Chen EWMA charts are often slightly less affected by the actual position of the change point than the ordinary Poisson EWMA chart. Because of this robustness regarding the time of change, the conclusions of the previous Section~\ref{Zero-state ARL Performance} still remain true for late changes. 

\smallskip
To sum up, if a sole mean shift happens under out-of-control conditions (no matter whether early or late), the ordinary Poisson EWMA chart is the superior choice. Otherwise, if we are also confronted with overdispersion, then an appropriate type of ABC-EWMA chart is the best solution. Good sensitivity towards both regular overdispersion or zero inflation is achieved by using $f(x)=|x-1|$ or $f(x)=\ln(x)$, where $|x-1|$ has slight advantages with respect to regular overdispersion, and $\ln(x)$ with respect to zero inflation. If we cannot be sure if a mean shift is combined with overdispersion or not, it is recommended to run both an ordinary and an ABC-EWMA chart in parallel.

\begin{figure}[t]
\small\centering
(a)\hspace{-3ex}%
\includegraphics[viewport=0 45 620 235, clip=, scale=0.5]{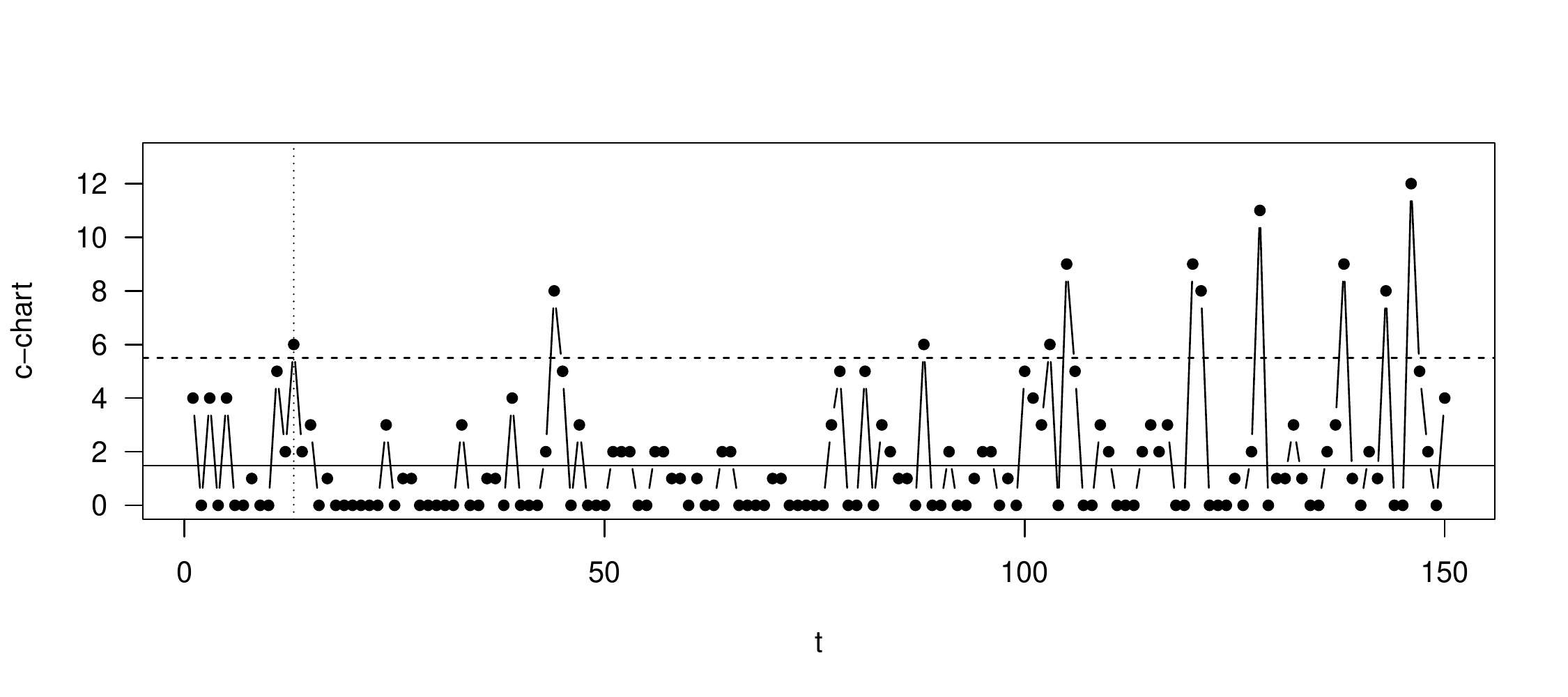}$t$
\\[2ex]
(b)\hspace{-3ex}%
\includegraphics[viewport=0 45 620 235, clip=, scale=0.5]{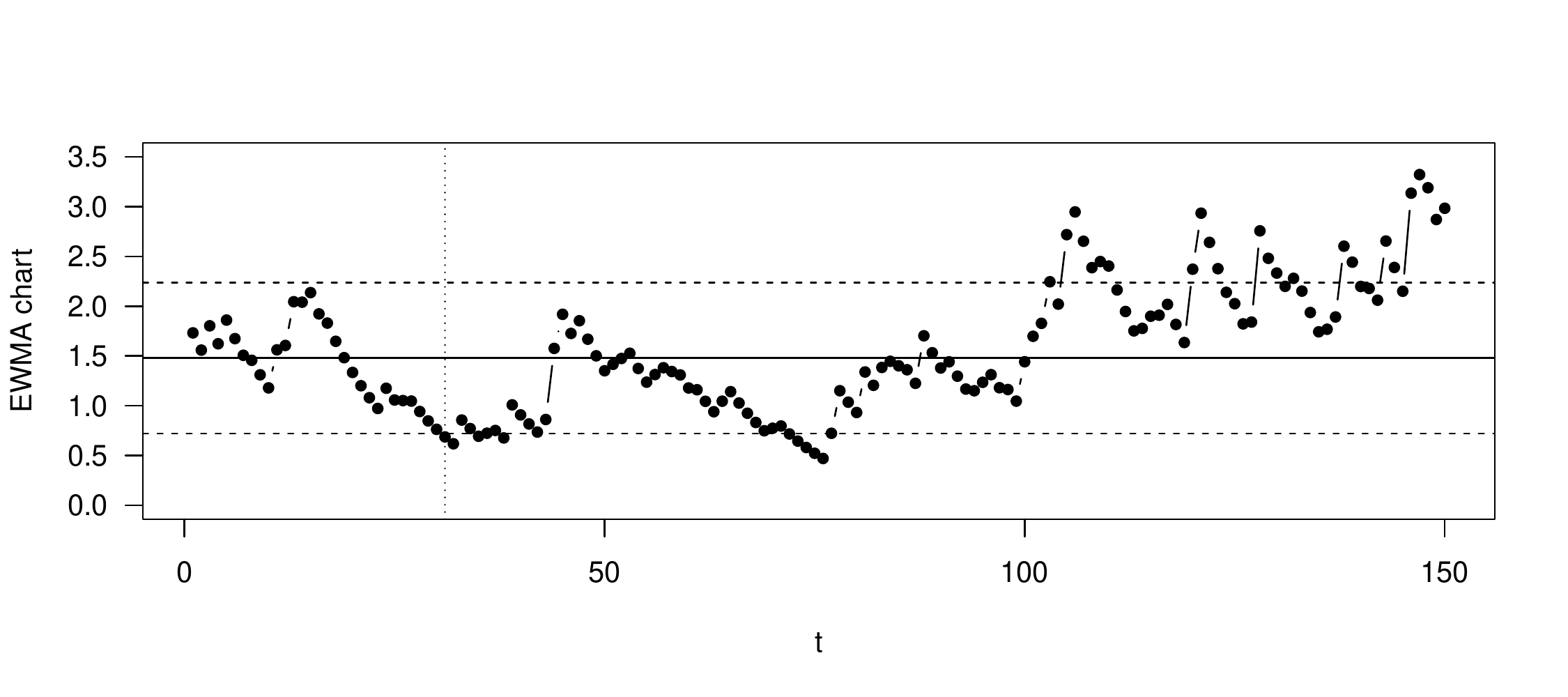}$t$
\caption{Phase-II analysis of particles data in Section~\ref{An Illustrative Data Example}: (a) c-chart and (b) ordinary EWMA chart. First alarm at dotted line.}
\label{figOrdinaryCharts}
\end{figure}

\section{An Illustrative Data Example}
\label{An Illustrative Data Example}
Let us analyze the particle counts data provided by \citet{nishina07}. Here, each count expresses the number of particles on a single wafer taken from a semiconductor manufacturing process. These data were already used by \citet{nishina07} to illustrate the effect of a possible model misspecification on process monitoring, namely if the data are erroneously modelled by a Poisson distribution. In the textbook literature, the Poisson distribution is the default choice for modeling unbounded counts and for designing corresponding control charts, see Section~7.3.1 in \citet{mont09}. But as we already know from Section~\ref{Simulation-based Performance Analyses}, an improper choice of the in-control model might severely deteriorate the ARL performance of the ordinary EWMA chart, so it is crucial to detect such kind of model misspecification as soon as possible. 

\smallskip
Thus, in line with \citet{nishina07}, where the particle counts data were used to illustrate the effects of an improper choice of the in-control model, we shall also design our control charts based on the erroneous assumption that the data are Poisson. More precisely, we split the original data set (200 observations) into two parts: we take the first $T_0=50$ observations for Phase-I analysis (model estimation and chart design under Poisson assumption), while the remaining 150 observations $x_1,\ldots,x_{150}$ are used for Phase-II analysis. Note that a plot of the data is provided by Figure~\ref{figOrdinaryCharts}\,(a), although the c-chart shown there also includes the center line and UCL (see the details below). The Phase-I data do not exhibit significant autocorrelations, \ie the data appear to be \iid, and their sample mean equals~1.48. As we are focussing on Poisson models like in the textbook literature, we finally conclude the in-control model $\poi(1.48)$. This is used to design the same types of control charts as considered in Section~\ref{Simulation-based Performance Analyses}, namely the
\begin{itemize}
	\item ordinary EWMA chart \eqref{ewmarecursion} with $L=0.758$ and ARL$_0\approx 370.9$;
	\item AB-EWMA chart \eqref{SCEWMA1b} with $f(x)=|x-1|$, $L=1.099$, and ARL$_0\approx 370.2$;
	\item AB-EWMA chart \eqref{SCEWMA1b} with $f(x)=|x-1|^{1/4}$, $L=0.986$, and ARL$_0\approx 371.3$;
	\item AB-EWMA chart \eqref{SCEWMA1b} with $f(x)=\ln(x)$, $L=1.017$, and ARL$_0\approx 371.3$;
	\item ABC-EWMA chart \eqref{SCEWMA2b} with $f(x)=|x-1|$, $L=0.638$, and ARL$_0\approx 368.8$;
	\item ABC-EWMA chart \eqref{SCEWMA2b} with $f(x)=|x-1|^{1/4}$, $L=0.574$, and ARL$_0\approx 369.5$;
	\item ABC-EWMA chart \eqref{SCEWMA2b} with $f(x)=\ln(x)$, $L=0.581$, and ARL$_0\approx 369.5$.
\end{itemize}
If trying to adapt a c-chart, then one cannot find a chart design with ARL$_0$ close to 370 (discreteness problem). With the (one-sided) decision rule that counts $\geq 6$ cause an alarm, we have ARL$_0\approx 239.2$; we shall take this chart design as a further competitor.

\begin{figure}[t]
\small\centering
(a)\hspace{-3ex}%
\includegraphics[viewport=0 45 620 235, clip=, scale=0.5]{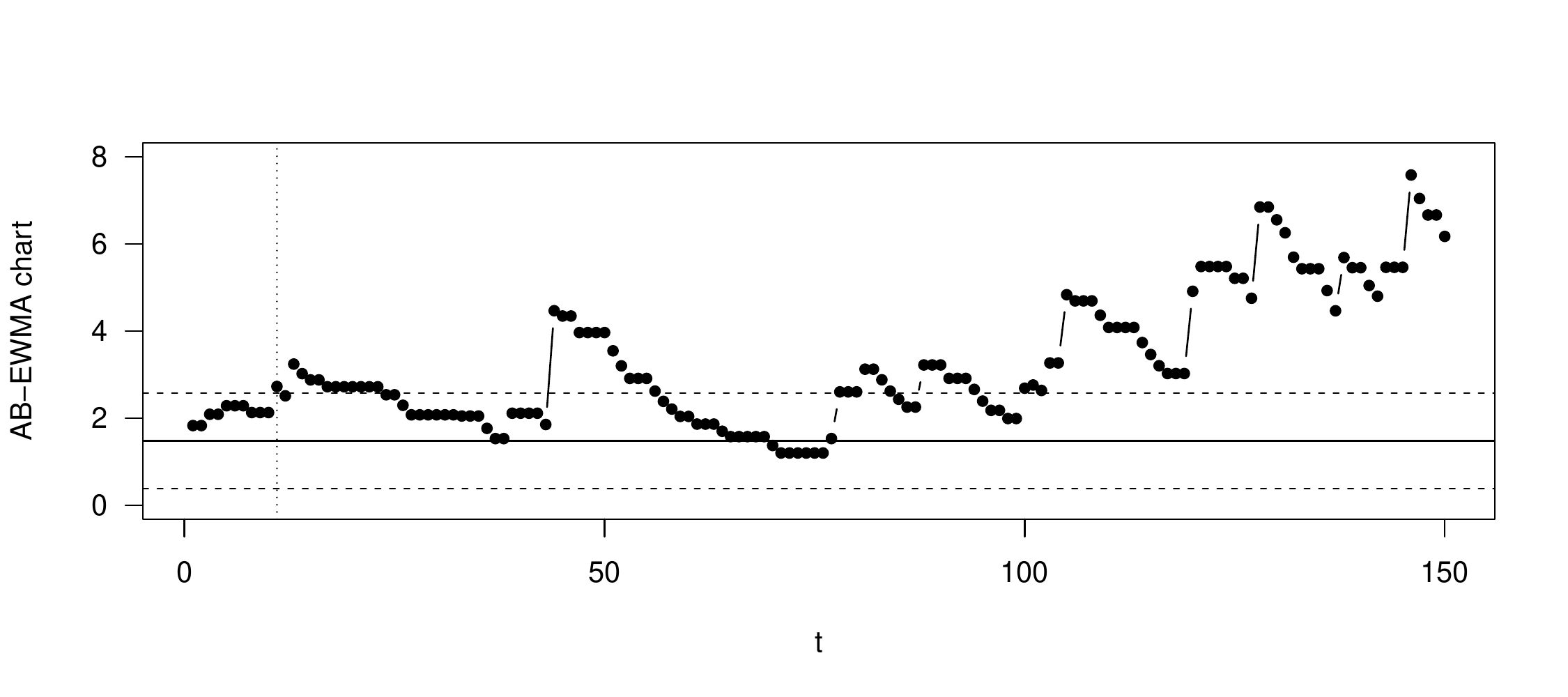}$t$
\\[2ex]
(b)\hspace{-3ex}%
\includegraphics[viewport=0 45 620 235, clip=, scale=0.5]{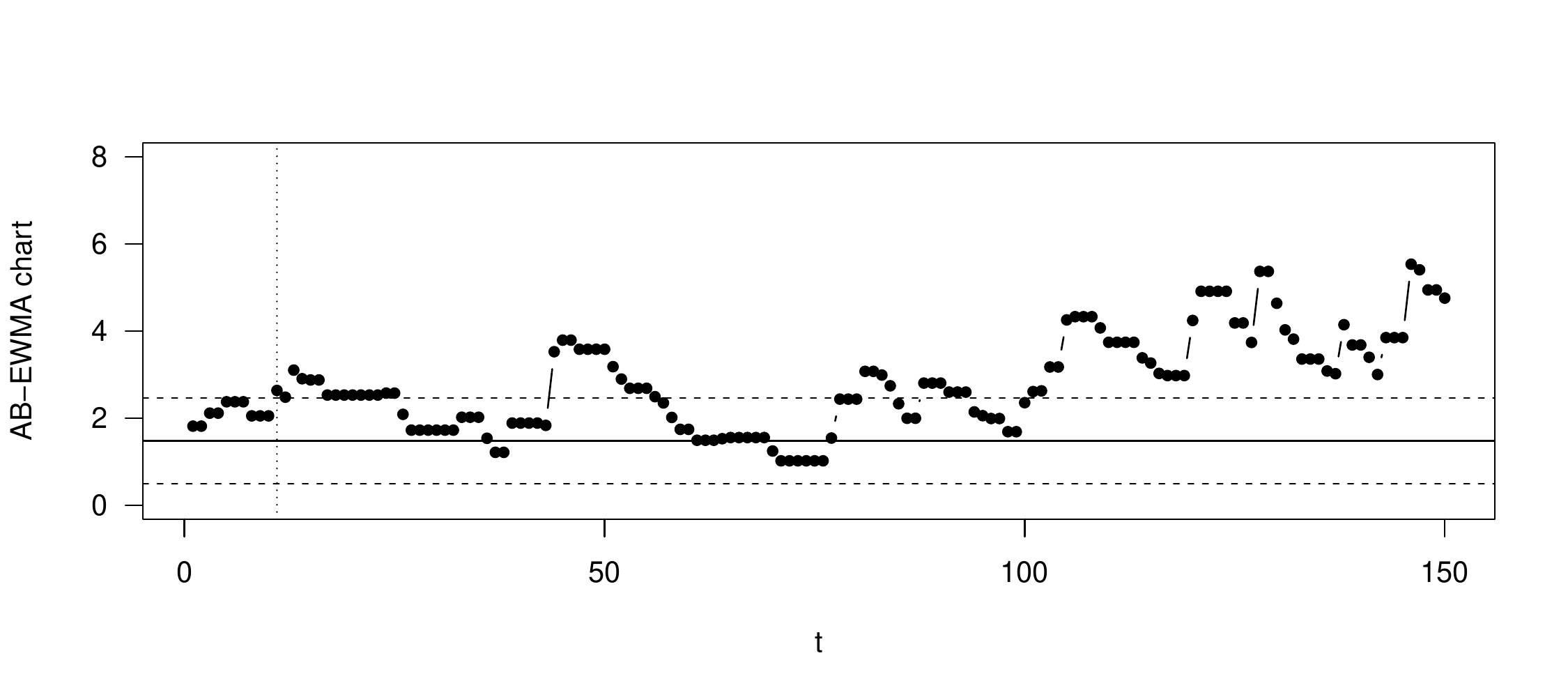}$t$
\\[2ex]
(c)\hspace{-3ex}%
\includegraphics[viewport=0 45 620 235, clip=, scale=0.5]{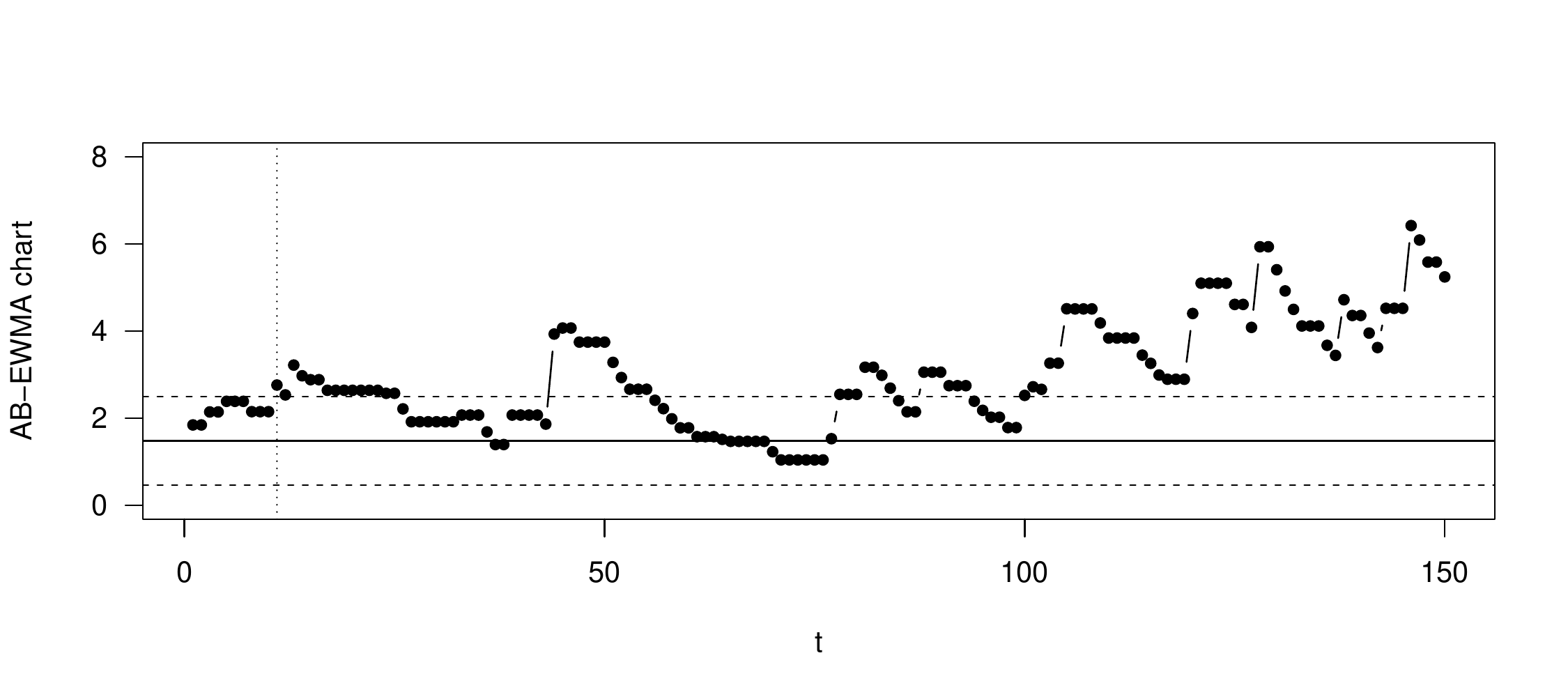}$t$
\caption{Phase-II analysis of particles data in Section~\ref{An Illustrative Data Example}: AB-EWMA charts with (a) $f(x)=|x-1|$, (b) $f(x)=|x-1|^{1/4}$, and (c) $f(x)=\ln(x)$. First alarm at dotted line.}
\label{figABEWMACharts}
\end{figure}

\begin{figure}[t]
\small\centering
(a)\hspace{-3ex}%
\includegraphics[viewport=0 45 620 235, clip=, scale=0.5]{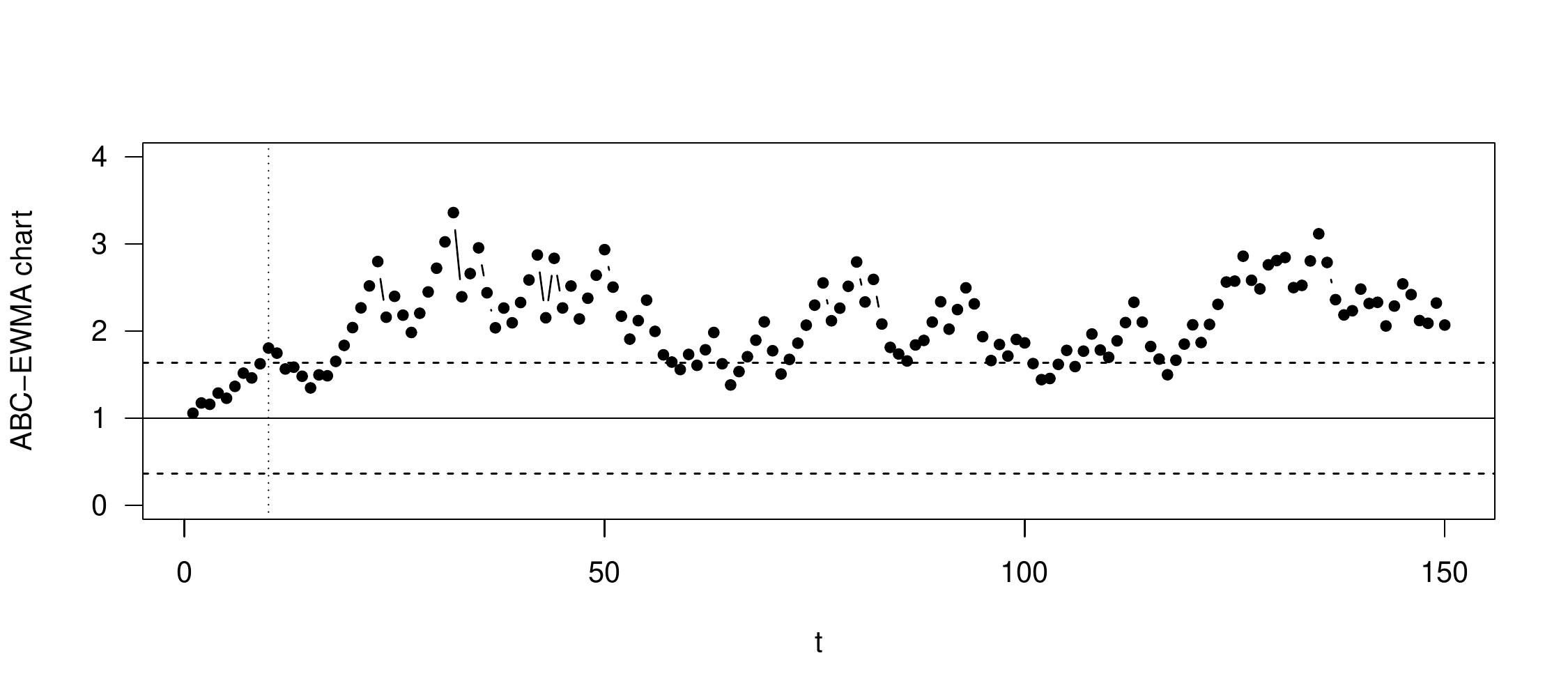}$t$
\\[2ex]
(b)\hspace{-3ex}%
\includegraphics[viewport=0 45 620 235, clip=, scale=0.5]{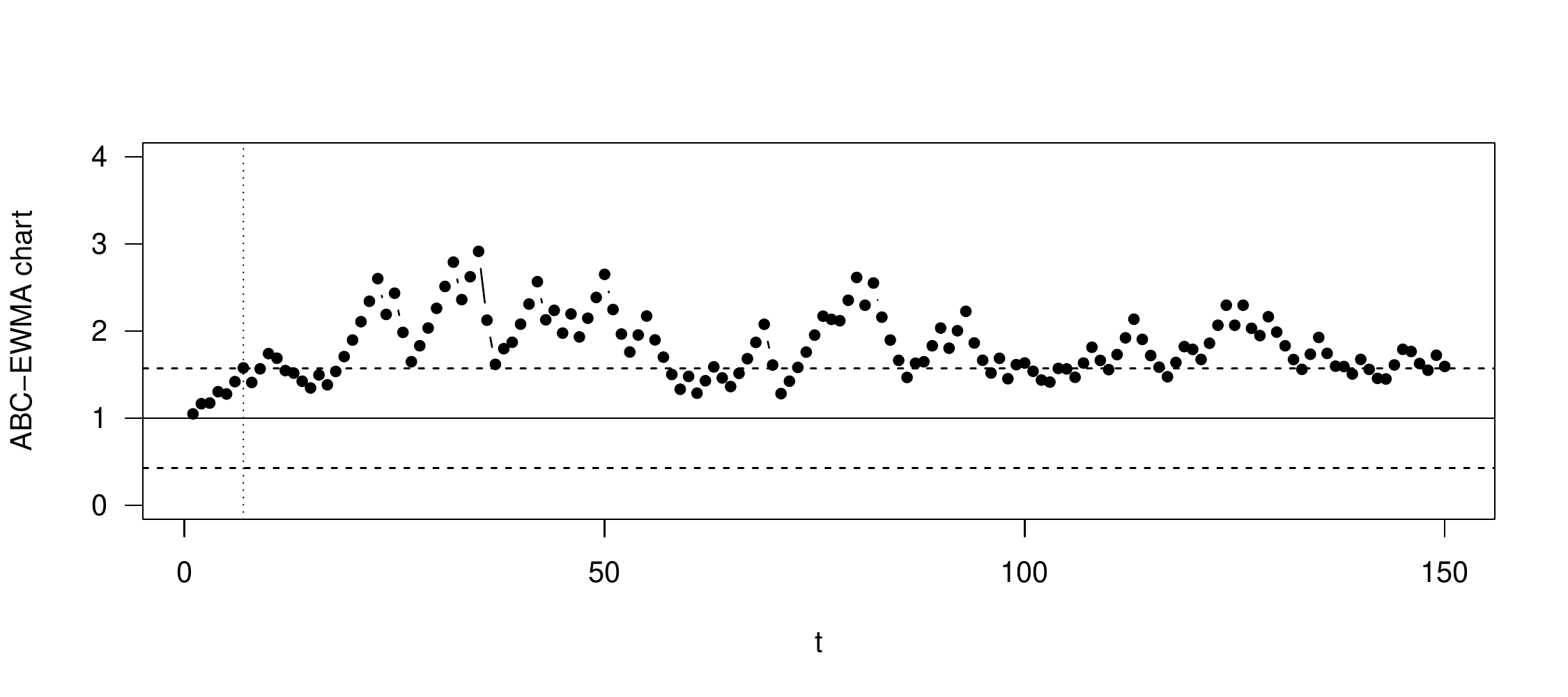}$t$
\\[2ex]
(c)\hspace{-3ex}%
\includegraphics[viewport=0 45 620 235, clip=, scale=0.5]{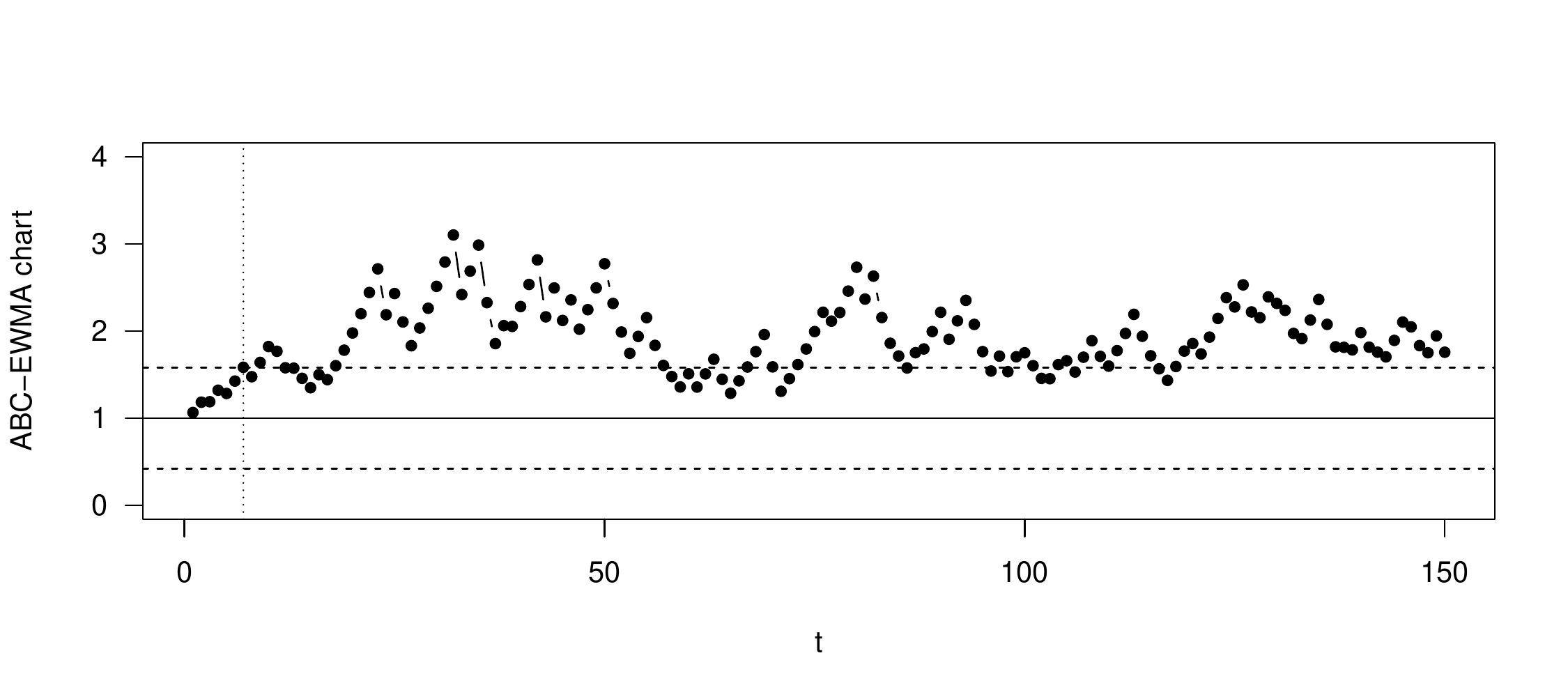}$t$
\caption{Phase-II analysis of particles data in Section~\ref{An Illustrative Data Example}: ABC-EWMA charts with (a) $f(x)=|x-1|$, (b) $f(x)=|x-1|^{1/4}$, and (c) $f(x)=\ln(x)$. First alarm at dotted line.}
\label{figABCEWMACharts}
\end{figure}

\medskip
Figures~\ref{figOrdinaryCharts}--\ref{figABCEWMACharts} show the results if applying the aforementioned control charts to the Phase-II data. All charts trigger several alarms, so the $\poi(1.48)$-model does not seem to be appropriate for these data. In fact, their sample mean 1.667 is slightly larger than $\mu_0=1.48$, but in particular, the data are strongly overdispersed, with sample dispersion index 3.525 (for this reason, \citet{nishina07} concluded that an NB-distribution would be more appropriate for the data). Thus, for practice, the crucial question is: Which chart is quickest in uncovering this obvious out-of-control situation? As can be seen from Figure~\ref{figOrdinaryCharts}, the ordinary EWMA chart is rather slow, with the first alarm triggered at time $t=31$. Even the simple c-chart is much faster, with the first alarm at $t=13$ (but also having a lower ARL$_0$ than the other charts). The slow reaction of the EWMA chart is explained by the fact that there is an only mild change in the mean, while its abilities in detecting overdispersion are quite limited, recall Section~\ref{Simulation-based Performance Analyses}.  

\smallskip
So let us now turn to the novel Stein--Chen EWMA charts. The AB-EWMA charts in Figure~\ref{figABEWMACharts} uniquely lead to a first alarm at $t=11$. Thus, they are much faster than the ordinary EWMA chart, and slightly faster than the c-chart. The superiority of the AB-EWMA charts is again plausible from our simulation results in Section~\ref{Simulation-based Performance Analyses}, although the latter give reason to hope that the ABC-EWMA charts do even better. In fact, see Figure~\ref{figABCEWMACharts}, any of these charts is faster in triggering its first alarm, namely the one with $f(x)=|x-1|$ at $t=10$, and those with $f(x)=|x-1|^{1/4}$ and $f(x)=\ln(x)$ at $t=7$. 
Altogether, we quickly recognize that the Poisson in-control model is not adequate for the particles counts data, and that some type of overdispersion appears to be present in the data. In fact, if we would have done the Phase-I analysis in a more comprehensive way, this would have got clear. The sample dispersion index of the Phase-I data equals $\hat{I}\approx 2.627$ and is thus considerably larger than one. In particular, if using~$\hat{I}$ to test $H_0: I=1$ (equidispersion) against $H_1: I>1$ (overdispersion), where $(T_0-1)\,\hat{I}$ is approximately $\chi_{T_0-1}^2$-distributed under the Poi-null \citep[see][4.7.4]{johnson05}, we get the P-value $4.416\cdot 10^{-9}$, which leads us to reject~$H_0$ in favour of the overdispersed alternative. So in practice, the next step would be to identify a new in-control model for the particle counts data.

\begin{figure}[t]
\includegraphics[viewport=0 45 260 235, clip=, scale=0.6]{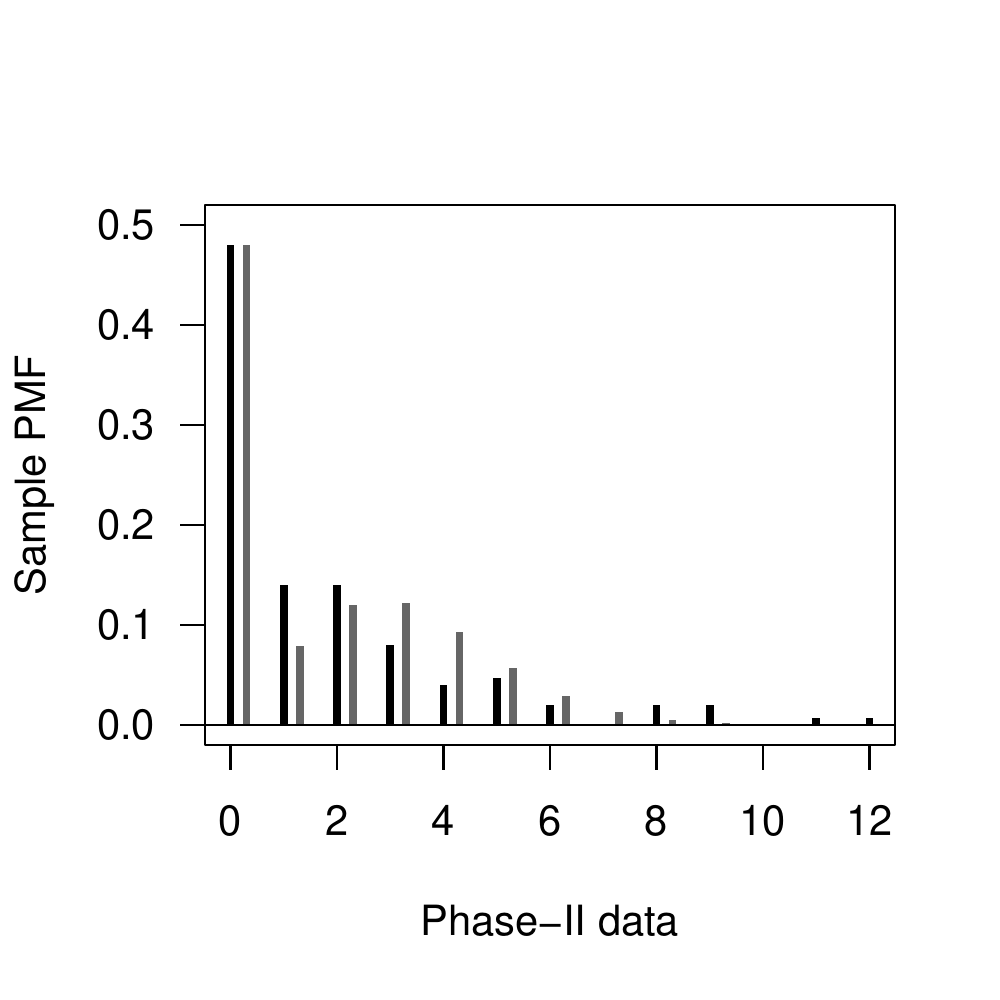}~$x$
\caption{Phase-II analysis of particles data in Section~\ref{An Illustrative Data Example}: plot of sample PMF (black bars) and plot of ML-fitted ZIP-PMF (grey bars).}
\label{figPhase2PMF}
\end{figure}

\smallskip
Let us conclude this application with a final remark. It is interesting to note that the ABC-EWMA charts with particular sensitivity regarding zero inflation were fastest. One might thus conjecture that the overdispersion is caused by zero inflation. In fact, computing the sample PMF of the Phase-II data, see the black bars in Figure~\ref{figPhase2PMF}, we observe an excessively large zero frequency, namely 48\,\%. Thus, a ZIP distribution appears to be a reasonable candidate for modeling the data (rather than the NB-distribution suggested by \citet{nishina07}). Fitting a ZIP distribution to the data (by maximum likelihood (ML) estimation, see the grey bars in Figure~\ref{figPhase2PMF}), we achieve a reasonable agreement. But we also recognize that counts $\geq 8$ occur too often. This agrees with the visual impression from Figure~\ref{figOrdinaryCharts}\,(a), where after $t=100$, such large counts are observed exceptionally often. Thus, it seems that we are not only concerned with overdispersion (caused by zero inflation), but also with a further mean shift after about $t=100$.

\section{Conclusions and Future Research}
\label{Conclusions}
In this article, two approaches were proposed of how to utilize the Stein--Chen identity when constructing an EWMA chart for Poisson counts; we referred to these charts as AB-EWMA and ABC-EWMA, respectively. While the ordinary EWMA chart is the best solution if sole mean shifts happen under out-of-control conditions, the novel Stein--Chen EWMA charts have appealing ARL properties if also (or only) changes in the distribution family happen (in the present research, the focus was on types of overdispersion), regardless of whether these changes occur early or late. More precisely, the concept of an ABC-EWMA chart (with an appropriate choice of the weight function~$f$) can clearly be recommended in such a case, whereas an AB-EWMA chart offers only limited advantages. The conclusions drawn from the simulation study were also confirmed by the real-data application on particle counts from a semiconductor manufacturing process. 

\medskip
There are several directions for future research. First, it is recommended to do more comprehensive performance analyses, also covering further out-of-control scenarios such as underdispersion or non-Poisson equidispersion. 
Second, Stein identities are available also for other count distributions than univariate Poisson \citep[see][]{sudheesh12,betsch22,weissaleksandrov22}, so the proposed Stein-type (ABC-)EWMA charts could be adapted to such in-control count distributions as well. This would be relevant, \eg for the particles count data, where a non-Poisson in-control model appears adequate. 
Third, Stein identities also exist for continuously distributed random variables \citep[see][]{stein81,sudheesh09,landsman16} and would thus allow to construct novel variables control charts. 
Fourth, for any of these Stein-type EWMA charts, it would be interesting to analyze the effect of parameter estimation (as done during Phase~I when identifying the in-control model) on the charts' performances, in analogy to the study by \citet{testik06} for the case of the ordinary Poisson EWMA chart. 
Finally, it would also be relevant to consider additional autocorrelation and to apply the Stein-type EWMA charts to time series data, see \citet{schmid97,knoth04,weiss15} for existing approaches. The potential of the Stein--Chen approach for time series data was recently demonstrated by \citet{aleksandrov22}, who developed GoF tests for Poisson count time series.

\subsubsection*{Acknowledgments}
The author thanks the referee for useful comments on an earlier draft of this article. 
This research was funded by the Deutsche Forschungsgemeinschaft (DFG, German Research Foundation) -- Projektnummer 437270842.

\end{document}